\documentclass[a4paper,10pt,twoside,twocolumn,fleqn]{article}
\usepackage[utf8]{inputenc}
\usepackage[T1]{fontenc}

\usepackage{fourier}

\usepackage{amsmath}
\allowdisplaybreaks
\usepackage{amsfonts}
\usepackage{bm}
\usepackage{mathrsfs}
\usepackage{booktabs}
\usepackage{graphicx}
\usepackage{url}

\usepackage{color,xcolor}

\usepackage{amsthm}
\theoremstyle{remark}
\newtheorem*{remark}{Remark}

\usepackage[english]{babel}

\usepackage{authblk}

\title{On the thermodynamic consistency of Quasi-Linear Viscoelastic models for soft solids}

\author[1]{Harold Berjamin}
\author[1]{Michel Destrade}
\author[2]{William J. Parnell}

\affil[1]{School of Mathematics, Statistics and Applied Mathematics, NUI Galway, University Road, Galway, Republic of Ireland}
\affil[2]{Department of Mathematics, University of Manchester, Oxford Road, Manchester M13 9PL, UK}

\date{}

\begin{document}
	
\twocolumn[
\begin{@twocolumnfalse}
	
\maketitle

\begin{abstract} \noindent
	Originating in the field of biomechanics, Fung's model of quasi-linear viscoelasticity (QLV) is one of the most popular constitutive theories employed to compute the time-dependent relationship between stress and deformation in soft solids. It is one of the simplest models of nonlinear viscoelasticity, based on a time-domain integral formulation. In the present study, we consider the QLV model incorporating a single scalar relaxation function. We provide natural internal variables of state, as well as a consistent expression of the free energy to illustrate the thermodynamic consistency of this version of the QLV model. The thermodynamic formulation highlights striking similarities between QLV and the internal-variable models introduced by Holzapfel and Simo. Finally, the dissipative features of compressible QLV materials are illustrated in simple tension. \\
	
\end{abstract}

\end{@twocolumnfalse}
]

\section{Introduction}\label{sec:Intro}

Nonlinear viscoelastic behaviour is observed in many soft solids, such as elastomers, gels, and biological materials. More specifically, the nonlinear mechanical response of viscous soft solids exhibits relaxation and creep phenomena in large deformation quasi-static tests. In dynamic tests, the response of such materials is sensitive to strain rates. Moreover, marked hysteresis loops in loading-unloading experiments are evident in such media \cite{holzapfel00, lemaitre09}.

Historically, the mechanical modelling of nonlinear viscoelastic solids has been approached in many different ways. Fundamentally, however, modelling viscoelastic effects amounts to a specification of the constitutive law to provide an accurate functional relationship between the instantaneous stress and the entire strain history \cite{drapaca07, wineman09}. In the monograph by Truesdell and Noll \cite{truesdell65}, this definition corresponds to the concept of \emph{simple materials} (Sec.~29 therein), see also Sec.~6.7 of the book by Malvern \cite{malvern69}.

If the stress depends only on a very short interval of the recent history of the deformation, then it can be expressed as a function of the time derivatives of the deformation gradient up to a finite order (cf. Truesdell and Noll \cite{truesdell65} Sec.~35). Often, time derivatives of the strain up to first order are considered (i.e., the stress is expressed in terms of strain and strain rates), which leads to Newtonian-type viscosity models \cite{destrade13,shariff17}. Similarly to the linear Kelvin--Voigt model, nonlinear \emph{strain-rate differential models} fail to describe stress relaxation phenomena \cite{banks11, carcione15}.

Differential models involving time derivatives of \emph{stress} as well as strain provide a natural bridge to models involving longer durations of time history. Indeed, the stress can then be expressed via an integral representation or via models with memory variables. These models can be seen as more or less equivalent, with memory variables manifesting themselves as terms within a stress relaxation function that arises in integral models. There is a vast literature for both integral and memory variable approaches. Reviews of viscoelastic constitutive modelling contrast these different approaches \cite{drapaca07, wineman09, holzapfel00}, or introduce them in a disconnected way \cite{lemaitre09, derooij16}.

\begin{figure}
	\centering
	\includegraphics{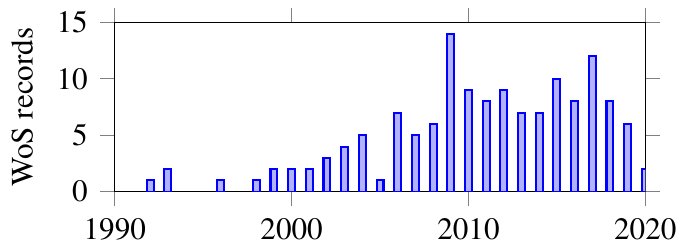}
	
	\caption{Number of QLV records by year in the Web of Science (WoS) database for the very specific query: \texttt{AB=(visco* AND quasi* AND (Fung OR QLV*))}.\label{fig:WoS}}
\end{figure}

A popular and simple model of the integral type is Fung's \emph{quasi-linear viscoelasticity} (QLV) \cite{fung81, depascalis14}, see Fig~\ref{fig:WoS}, which is based on a Boltzmann superposition principle. The instantaneous stress is expressed as a convolution product between a relaxation tensor and the elastic stress response. Originating in biomechanics, Fung's QLV has been successfully employed in related applications, with the development of associated experimental techniques \cite{sarver03,rashid13,babaei15} and dedicated computational methods \cite{puso98,holzapfel17}. It has also been employed to model polymers and rubbers \cite{asif19, jridi18}.

The computation of the instantaneous QLV stress by means of the convolution product requires the storage of the whole strain history, since the constitutive law is non-local in time. In computational applications, this can be avoided by expanding the relaxation tensor as a Prony series, leading to the natural definition of memory variables \cite{puso98, holzapfel17}. Suitable memory variables account for the deformation history in a time-local fashion \cite{taylor09}, allowing efficient evaluation of the mechanical response.

As described extensively by Maugin \cite{maugin99, maugin15}, constitutive models with memory variables rely on \emph{thermodynamics with internal variables of state} to ensure that the addition of new variables entails a dissipative contribution. Various viscoelastic constitutive models of the literature satisfy these principles by design \cite{holzapfel00}. In particular, viscous strain variables may be consistently introduced within a multiplicative decomposition of the deformation gradient tensor, in combination with Maxwell-type rheologies \cite{lubliner85, lejeunes11}.

To the present authors' knowledge, a thermodynamically-consistent expression of the energy in terms of the memory variables of QLV is yet to be presented.
In this paper, we recall the equations governing the motion of QLV solids with a single scalar relaxation function (Sec.~\ref{sec:GovEq}). The main result is derived in Sec.~\ref{sec:Thermo}, namely a thermodynamically consistent expression of the free energy with appropriate internal variables of state. This way, we establish links between Fung's QLV and the internal-variable models by Simo \cite{simo87}, and Holzapfel and Simo \cite{holzapfel96b}. Using a neo-Hookean model, the dissipative features of compressible QLV solids are illustrated in Sec.~\ref{sec:Illustrations}. The incompressible case is addressed in \ref{app:Incomp}.

\section{Governing equations}\label{sec:GovEq}

\subsection{Preliminaries}\label{ssec:GovEqGen}

In what follows, we present the basic equations of Lagrangian solid dynamics \cite{holzapfel00, ogden84}. We consider a homogeneous and isotropic solid continuum on which no external volume force is applied. A particle initially  at position $\bm{X}$ in the reference configuration moves to position $\bm{x}$ in the current configuration. The deformation gradient tensor is the second-order tensor
\begin{equation}
\bm F = \frac{\partial{\bm{x}}}{\partial{\bm{X}}} = \bm{I} + \text{Grad}\, \bm{u} \, ,
\label{F}
\end{equation}
where $\bm{u} = \bm{x} - \bm{X}$ is the displacement field, $\bm{I}$ is the identity tensor, and $\text{Grad}$ denotes the gradient operator with respect to the material coordinates ${\bm{X}}$ (Lagrangian gradient). If the Euclidean space is described by an orthonormal basis $\lbrace \bm{e}_1, \bm{e}_2, \bm{e}_3 \rbrace$ and a Cartesian coordinate system, then $\bm{I} = [\delta_{ij}]$, where $\delta_{ij}$ is the Kronecker delta.
The volume dilatation
\begin{equation}
	J = \det\bm{F}
	\label{Dilat}
\end{equation}
equals the ratio ${\rho_0}/{\rho}$ of the mass densities in the reference (undeformed) and deformed configurations.

One can define various strain tensors as functions of $\bm F$, such as the right Cauchy--Green deformation tensor $\bm{C} = \bm{F}^\top\! \bm{F}$, and the Green--Lagrange strain tensor $\bm{E} = \frac{1}{2} (\bm{C} - \bm{I})$. Frequently, principal stretches $\lambda_i$ are introduced, whose squares $\lambda_i^2$ correspond to the eigenvalues of $\bm{C}$. Thus, the principal invariants $I_i$ of $\bm C$ are given by
\begin{equation}
	\begin{aligned}
		& I_1 = \text{tr}\, \bm{C} = \lambda_1^2 + \lambda_2^2 + \lambda_3^2 \\
		& I_2 = \tfrac12 \big((\text{tr}\, \bm{C})^2 - \text{tr} (\bm{C}^2)\big) = \lambda_1^2\lambda_2^2 + \lambda_2^2\lambda_3^2 + \lambda_1^2\lambda_3^2 \\
		& I_3 = \det \bm{C} = \lambda_1^2 \lambda_2^2 \lambda_3^2 = J^2 .
	\end{aligned}
	\label{Invariants}
\end{equation}

The dynamics of the continuum in question are governed by conservation of momentum, which involves the divergence of a stress tensor. 
Typically, in the Lagrangian description, this equation of motion  involves the first Piola--Kirchhoff stress tensor $\bm P$, and the Eulerian version involves the Cauchy stress tensor $\bm{\sigma} = J^{-1}\bm{P} \bm{F}^\top\!$. Specified by the constitutive law, these stress measures  are also related to the second Piola--Kirchhoff stress tensor $\bm{S} = \bm{F}^{-1} \bm{P}$.

\subsection{Fung's quasi-linear viscoelasticity}\label{ssec:FungModel}

Fung's \emph{quasi-linear viscoelasticity} (QLV) is presented below (see Sec.~7.13 of \cite{fung81}). This model is based on the assumption that the stress is linearly dependent on the history of the elastic stress response, and a Boltzmann superposition principle between both quantities is assumed. The second Piola--Kirchhoff stress is given by
\begin{equation}
\bm{S} = \bm{\mathsf G} \bm{*} \dot{\bm S}^\text{e} = \int_{\mathbb R} \bm{\mathsf G}(t-s) : \dot{\bm S}^\text{e}(s) \, \text d s = \dot{\bm{\mathsf G}} \bm{*} {\bm S}^\text{e} \, ,
\label{FungConv}
\end{equation}
where the elastic response \cite{depascalis14}
\begin{equation}
\begin{aligned}
	\bm{S}^\text{e} &= \partial W/\partial\bm{E} = 2\, \partial W/\partial\bm{C} \\
	 &= 2 \left(W_1 + I_1 W_2\right)\bm{I} - 2 W_2 \bm{C} + 2I_3 W_3 \bm{C}^{-1} \\
	&= 2 W_1 \bm{I} + 2 \left(I_2W_2 + I_3 W_3\right) \bm{C}^{-1} - 2 I_3 W_2 \bm{C}^{-2}
\end{aligned}
\label{Elastic}
\end{equation}
is derived from a strain energy density function $W(I_1, I_2, I_3)$, and $\bm{\mathsf G}$ is a fourth-order relaxation tensor. Here the colon denotes the double contraction $\bm{\mathsf G} : \dot{\bm S}^\text{e} = [\bm{\mathsf G}_{ijk\ell} {\dot{\bm S}^\text{e}}_{k\ell}]$ where Einstein notation is used and the dot denotes the material time derivative. The notation $W_i$ is shorthand for the derivative $\partial W/\partial I_i$.

Also, in a similar fashion to Taylor et al. \cite{taylor09}, we introduce the Flory decomposition of the deformation into volumetric and deviatoric parts (see Holzapfel \cite{holzapfel00} Sec. 6.4). Thus, we introduce the volume-preserving Cauchy--Green strain tensor $\tilde{\bm C} = J^{-2/3} \bm{C}$ and its volume-changing counterpart $J^{2/3}\bm{I}$.
We perform the change of variable $W = \tilde W(\tilde I_1, \tilde I_2, J)$ in the expression of the strain energy, where
\begin{equation}
	\tilde I_1 = J^{-2/3}I_1\, ,\quad
	\tilde I_2 = J^{-4/3} I_2\, ,\quad
	J = \sqrt{I_3} \; .
	\label{FloryInvariants}
\end{equation}
The invariants $\tilde I_i$ in Eq.~\eqref{FloryInvariants} describe volume-preserving deformation, while the dilatation $J$ describes volume-changing deformation. The third invariant $\tilde I_3$ of $\tilde{\bm C}$ is equal to one, as deduced from the definitions in Eq.~\eqref{Invariants}.

To compute the elastic response \eqref{Elastic} in terms of the new variables $(\tilde I_1, \tilde I_2, J)$, let us recall expressions for the tensor derivatives $\partial\tilde{\bm C}/\partial\bm{C} = J^{-2/3} \big(\bm{\mathsf I} - \frac13 \bm{C}\otimes\bm{C}^{-1}\big)$ and $\partial J/\partial \bm{C} = \frac12 J \bm{C}^{-1}\!$, where the fourth-order unit tensor is defined as $\bm{\mathsf I} = [\delta_{ik}\delta_{j\ell}]$. Thus, we introduce the decomposition
\begin{equation}
\begin{aligned}
	\bm{S}^\text{e} &= \bm{S}^\text{e}_\text{D} + \bm{S}^\text{e}_\text{H} \, ,\\
	\bm{S}^\text{e}_\text{D} &= J^{-2/3}\text{Dev}(\tilde{\bm S}^\text{e})
	\quad\text{with}\quad
	\tilde{\bm S}^\text{e} = 2\, {\partial\tilde W}/{\partial\tilde{\bm C}} \, ,
	\\
	\bm{S}^\text{e}_\text{H} &= ({\partial\tilde W}/{\partial J})\, J\bm{C}^{-1} ,
\end{aligned}
\label{ElasticSplitTilde}
\end{equation}
where $\text{Dev}(\bullet) = (\bullet) - \frac13 (\bullet : \bm{C})\bm{C}^{-1}\!$ denotes the deviatoric operator in the Lagrangian description \cite{holzapfel00}.
The expression of $\tilde{\bm S}^\text{e}$ is deduced from Eq.~\eqref{Elastic} with $\tilde W_3 = 0$.
Converting back to the variables $(I_1, I_2, I_3)$, the chain rule yields
\begin{equation}
	\begin{aligned}
		\bm{S}^\text{e}_\text{D} &= 2(W_1 + I_1 W_2)\bm{I} - 2 W_2\bm{C} - \tfrac23 \left(I_1 W_1 + 2 I_2 W_2\right) \bm{C}^{-1} \\
		&= 2W_1\bm{I} + \tfrac23 \left(I_2 W_2 - I_1 W_1\right) \bm{C}^{-1} - 2I_3W_2\bm{C}^{-2} ,\\
		\bm{S}^\text{e}_\text{H} &= \tfrac23 \left(I_1 W_1 + 2 I_2 W_2 + 3 I_3 W_3\right) \bm{C}^{-1} ,
	\end{aligned}
	\label{ElasticSplit}
\end{equation}
which are the same expressions as in De Pascalis et al. \cite{depascalis14} (Eqs.~(3.19)-(3.20) therein). In a standard fashion \cite{holzapfel00}, Taylor et al. \cite{taylor09} assumes the separability of isochoric and volumetric deformations $W = \tilde W^\text{iso}(\tilde I_1, \tilde I_2) + \tilde W^\text{vol}(J)$, which is a particular case of the present expressions.

In a similar fashion to related works \cite{drapaca07,taylor09}, we assume that the relaxation is the same in all directions, i.e. $\bm{\mathsf G} = \mathscr{G}\bm{\mathsf I}^\text{s}$ where we have defined the fourth-order symmetric identity tensor $\bm{\mathsf I}^\text{s} = \frac12 [\delta_{ik}\delta_{j\ell} + \delta_{i\ell}\delta_{jk}]$, and $\mathscr{G}$ is a scalar function. Thus, the constitutive law \eqref{FungConv} yields 
\begin{equation}
	\bm{S} = \mathscr{G} * \dot{\bm S}^\text{e} = \dot{\mathscr{G}} * {\bm S}^\text{e} .
	\label{FungConvScalar}
\end{equation}
Fung  \cite{fung81} initially proposed an integral expression of $\mathscr{G}$ with a continuous spectrum of relaxation. This expression leads to high computational costs as it requires the storage of the whole deformation history \cite{banks11}. This drawback can be avoided by approximating $\mathscr{G}$ as an exponential series \cite{puso98, holzapfel17}\footnote{The increasing exponentials in cited literature indicate a potential error.} of the form
\begin{equation}
	\mathscr{G}(t) = \bigg(1 - \sum_{k=1}^n g_k\, (1-\text{e}^{-t/\tau_k})\bigg) \operatorname{H}(t) \, ,
	\label{PronyRel}
\end{equation}
with an arbitrary number $n$ of relaxation mechanisms \cite{taylor09, rashid13}. The Heaviside step function $\operatorname{H}$ is included in Eq.~\eqref{PronyRel} for convenience. Such a Prony series with magnitudes $g_k>0$ and characteristic relaxation times $\tau_k>0$ can be linked to generalised Maxwell-type rheologies.

\section{Thermodynamics}\label{sec:Thermo}

\subsection{Generalities}\label{ssec:ThermoGen}

The consequences of the first and second principles of thermodynamics are summarized below. We consider deformable solids whose associated constitutive law involves $n$ second-order tensorial internal variables of state $\bm{\alpha}_1, \dots, \bm{\alpha}_n$ \cite{maugin99,maugin15}.

\paragraph{Isentropic modelling} We consider the set of variables of state $\lbrace \eta , \bm{E} , \bm{\alpha}_1, \dots, \bm{\alpha}_n\rbrace$, where $\eta$ denotes the entropy per unit mass. The state is assumed local in time, i.e., only its instantaneous value is considered. The first principle of thermodynamics is reflected in the conservation of energy $\rho\dot{e} = \bm{\sigma}:\bm{D}$, where $\dot{e}$ is the material time-derivative of the internal energy $e$ per unit mass, and $\bm D = \bm{F}^{-\top}\! \dot{\bm E} \bm{F}^{-1}\!$ denotes the strain-rate tensor (i.e., the symmetric part of the Eulerian velocity gradient $\dot{\bm{F}}\bm{F}^{-1}$). The second principle of thermodynamics imposes the increase of entropy $\rho \dot{\eta} \geq 0$. Assuming an adiabatic process, the dissipation per unit of reference volume reads $\mathscr{D} = \rho_0 T\dot{\eta}$ (W/m\textsuperscript{3}), where $T> 0$ is the absolute temperature. Thus, combining the local equations of thermodynamics with the Gibbs identity, the Clausius--Duhem inequality is obtained:
\begin{equation}
\begin{aligned}
\mathscr{D} &= \rho_0 \left(T - \frac{\partial e}{\partial \eta}\right) \dot{\eta} + J\bm{\sigma}:\bm{D} - \frac{\partial U}{\partial \bm{E}} : \dot{\bm E} - \sum_{k=1}^n \frac{\partial U}{\partial \bm{\alpha}_k} : \dot{\bm \alpha}_k \\ & \geq 0\, .
\end{aligned}
\label{CDS1}
\end{equation}
Here, $U = \rho_0 e$ is the internal energy per unit reference volume.
Since the inequality \eqref{CDS1} must be satisfied for all states and all evolutions, the coefficient $T - \partial e/\partial \eta$ must equal zero.
Using the identity $J\bm{\sigma} : \bm{D} = \bm{S}: \dot{\bm E}$, see e.g. Ref.~\cite{berjamin17}, the Clausius--Duhem inequality \eqref{CDS1} is rewritten as
\begin{equation}
\mathscr{D} = \left(\bm{S} - \frac{\partial U}{\partial \bm{E}}\right):\dot{\bm E} - \sum_{k=1}^n \frac{\partial U}{\partial \bm{\alpha}_k} : \dot{\bm \alpha}_k \geq 0\, .
\label{CDS2}
\end{equation}
We call this framework \textit{isentropic} because the partial derivatives ${\partial }/{\partial \bm{E}}$, ${\partial }/{\partial \bm{\alpha}_k}$ are evaluated at constant entropy \cite{holzapfel00}.

\paragraph{Isothermal modelling} This approach involves the variables of state $\lbrace T , \bm{E} , \bm{\alpha}_1, \dots, \bm{\alpha}_n\rbrace$. It is linked to the above expressions by introducing the partial Legendre transform $\psi = e-T\eta$ of $e$, which is Helmholtz' free energy per unit mass. We have
\begin{equation}
	\eta = -\frac{\partial \psi}{\partial T} \, ,
	\quad \frac{\partial U}{\partial \bm E} = \frac{\partial \Psi}{\partial \bm E} \, ,
	\quad \frac{\partial U}{\partial \bm{\alpha}_k} = \frac{\partial \Psi}{\partial \bm{\alpha}_k} \, ,
\end{equation}
where $\Psi = \rho_0 \psi$ is the Helmholtz free energy per unit of reference volume. The Clausius--Duhem inequality \eqref{CDS2} is re-expressed as
\begin{equation}
	\mathscr{D} = \left(\bm{S} - \frac{\partial \Psi}{\partial \bm{E}}\right):\dot{\bm E} - \sum_{k=1}^n \frac{\partial \Psi}{\partial \bm{\alpha}_k} : \dot{\bm \alpha}_k \geq 0\, .
	\label{CDT}
\end{equation}
We observe that thermodynamic restrictions have the same form in the isentropic and in the isothermal frameworks. In the isothermal framework described here, proving thermodynamic consistency of Fung's QLV \eqref{FungConvScalar} amounts to finding $\Psi$, $\bm{\alpha}_k$ such that the Clausius--Duhem inequality \eqref{CDT} is always satisfied.

\subsection{Fung's quasi-linear viscoelasticity}\label{ssec:ThermoFung}

\paragraph{Memory variables}

Using the expression of the relaxation function \eqref{PronyRel}, the constitutive law \eqref{FungConvScalar} is rewritten as \cite{taylor09}
\begin{equation}
\bm{S} = \bm{S}^\text{e} - \sum_{k=1}^n \bm{S}^\text{v}_k
	\label{FungMemStress}
\end{equation}
where the viscous stress
\begin{equation}
	\bm{S}^\text{v}_k = g_k \int_0^t \big(1-\text{e}^{-(t-s)/\tau_k}\big)\, \dot{\bm S}^\text{e}(s) \, \text d s
\label{FungMemVar}
\end{equation}
can thus be interpreted as a memory variable. Computing its material time-derivative, one shows that $\bm{S}^\text{v}_k$ satisfies the linear evolution equation \cite{taylor09}
\begin{equation}
\tau_k \dot{\bm S}^\text{v}_k = g_k \bm{S}^\text{e} - \bm{S}^\text{v}_k\, .
\label{FungEvol}
\end{equation}
Thus, the convolution product \eqref{FungConvScalar} is replaced by a sum of $n$ memory variables, which satisfy a linear differential equation.

By construction, Fung's QLV model reduces to hyperelasticity for particular relaxation functions involving certain limits of relaxation times:
\begin{itemize}
	
	\item \emph{Relaxed elastic solid.}
	The relaxed elastic limit corresponds to infinite durations, i.e. to short relaxation times $\tau_k \to 0$. Hence, the evolution equation \eqref{FungEvol} produces ${\bm{S}}_k^\text{v} = g_k {\bm{S}}^\text{e}$. If the motion is causal, the convolution product \eqref{FungConvScalar} reduces to $\bm{S} = (1 - \sum_{k} g_k)\, {\bm{S}}^\text{e}$ where the coefficient $1 - \sum_{k} g_k$ defines the relaxed elastic modulus.
	
	\item \emph{Unrelaxed elastic solid.}
	The unrelaxed elastic limit corresponds to infinitesimal durations, i.e. to long relaxation times $\tau_k \to +\infty$. Hence, the evolution equation \eqref{FungEvol} gives ${\bm{S}}_k^\text{v} = \bm{0}$ for causal signal. The convolution product \eqref{FungConvScalar} reduces to $\bm{S} = {\bm{S}}^\text{e}$. With respect to the relaxed elastic solid, the effective elastic moduli differ by a scalar coefficient.
	
\end{itemize}
These elastic limits correspond to zero dissipation \cite{parnell19}.

\paragraph{Dissipation}

Consider a (presumably convex) strain energy function $W$ from which the elastic response $\bm{S}^\text{e}$ is obtained by differentiation. We define the free energy in such a way that $\bm{S} = \partial \Psi/\partial \bm{E}$ is satisfied:
\begin{equation}
	\Psi = W(\bm{E}) - \sum_{k=1}^n \left(\bm{S}_k^\text{v} : \bm{E} - \Phi_k(\bm{S}_k^\text{v})\right)  .
	\label{PotPotStress}
\end{equation}
The arbitrary functions $\Phi_k$ are (presumably convex) potentials whose dependency on the variables $\bm{S}_k^\text{v}$ of Eq.~\eqref{FungMemVar} needs to be specified. 
If the viscous stresses $\bm{S}_k^\text{v}$ governed by Eq.~\eqref{FungEvol} are internal variables of state $\bm{\alpha}_k$, then the dissipation \eqref{CDT} reads
\begin{equation}
	\begin{aligned}
	\mathscr{D} &= -\! \sum_{k=1}^n \frac{\partial \Psi}{\partial \bm{S}_k^\text{v}} : \dot{\bm S}_k^\text{v} \\
	&= \sum_{k=1}^n \frac1{\tau_k} \left(\bm{E} - \frac{\partial \Phi_k}{\partial \bm{S}_k^\text{v}} \right) : \left(g_k \bm{S}^\text{e} - \bm{S}^\text{v}_k\right) .
	\end{aligned}
	\label{PotDissStress}
\end{equation}
Let us introduce the Legendre transform $W_k(\bm{E}_k^\text{v}) = \bm{S}_k^\text{v} : \bm{E}_k^\text{v} - \Phi_k(\bm{S}_k^\text{v})$
of $\Phi_k$ such that $\bm{E}_k^\text{v} = {\partial \Phi_k}/{\partial \bm{S}_k^\text{v}}$ is the conjugate variable of $\bm{S}_k^\text{v} = \partial W_k/\partial\bm{E}_k^\text{v}$. The free energy \eqref{PotPotStress} and the dissipation \eqref{PotDissStress} become
\begin{equation}
	\begin{aligned}
	\Psi & = W(\bm{E}) - \sum_{k=1}^n \left(W_k(\bm{E}_k^\text{v}) + \frac{\partial W_k}{\partial\bm{E}_k^\text{v}} : (\bm{E} - \bm{E}_k^\text{v}) \right) , \\
	\mathscr{D} &= \sum_{k=1}^n \frac{1}{\tau_k} \left(\bm{E} - \bm{E}_k^\text{v} \right) : \left(g_k\frac{\partial W}{\partial \bm{E}} - \frac{\partial W_k}{\partial \bm{E}_k^\text{v}}\right) .
	\end{aligned}
	\label{PotInternal}
\end{equation}
To ensure the positivity of the dissipation, a compatible choice of potentials is $W_k(\cdot) = g_k W(\cdot)$ pointwise. Consequently, the scaling property of the Legendre transformation imposes $\Phi_k(\cdot) = g_k \Phi(\cdot/g_k)$ pointwise, where the potential $\Phi$ known as \emph{complementary energy density} defines the Legendre transform
$\bm{S}^\text{e} : \bm{E} - W(\bm{E})$
of $W$, with $\bm{E} = \partial \Phi/\partial \bm{S}^\text{e}$ \cite{fung79,ogden84}.
Finally, we have
\begin{equation}
	\begin{aligned}
	\Psi &= W(\bm{E}) - \sum_{k=1}^n g_k \left(W(\bm{E}_k^\text{v}) + \frac{\partial W(\bm{E}_k^\text{v})}{\partial \bm{E}_k^\text{v}}:(\bm{E} - \bm{E}_k^\text{v})\right) ,\\
	\mathscr{D} &= \sum_{k=1}^n \frac{g_k}{\tau_k} \left(\bm{E} - \bm{E}_k^\text{v} \right) : \left(\frac{\partial W(\bm{E})}{\partial \bm{E}} - \frac{\partial W(\bm{E}_k^\text{v})}{\partial \bm{E}_k^\text{v}}\right) .
	\end{aligned}
	\label{PotInternalFinal}
\end{equation}
By virtue of the convexity inequality \cite{ogden84}, the dissipation $\mathscr D$ is non-negative for any convex strain energy density function $W$ of $\bm E$. Therefore, the present compressible QLV model is thermodynamically admissible.

\paragraph{Connections with other models} The derivation of the model introduced by Simo \cite{simo87} is very similar. In fact, this model is based on the free energy
\begin{equation}
	\Psi = W(\bm{E}) - \sum_{k=1}^n \left(\bm{S}_k^\text{v} : \tilde{\bm E} - \Phi_k(\bm{S}_k^\text{v})\right) ,
 	\label{Simo}
\end{equation}
where the separability of isochoric and volumetric deformations $W = \tilde W^\text{iso}(\tilde I_1, \tilde I_2) + \tilde W^\text{vol}(J)$ is assumed. Similarly to Eq.~\eqref{FungEvol}, a consistent linear evolution equation for the viscous stresses is proposed. Thus, the main difference between Eqs.~\eqref{Simo} and \eqref{PotPotStress} lies in the presence of the volume-preserving version $\tilde{\bm E} = \frac12 (\tilde{\bm C} - \bm{I})$ of $\bm E$. Note that the ``Simo'' model described in the MSC Nastran Implicit Nonlinear user guide is actually a QLV model (Ref.~\cite{nastran}, Chap.~10, Eq.~(10-88)). Nevertheless, that implementation assumes ``that the viscoelastic behavior \textellipsis acts only on the deviatoric behavior''. While it remains unclear how this is done, the implementation may therefore be consistent with the original Simo model \cite{simo87}.

The free energy \eqref{PotPotStress} of Fung's QLV can be rewritten as $\Psi = \Psi^\infty + \sum_{k} \Upsilon_k$ with
\begin{equation}
	\begin{aligned}
	\Psi^\infty &= \bigg(1 - \sum_{k=1}^n g_k\bigg)\, W(\bm{E}) \, ,
	\\
	\Upsilon_k &= g_k W(\bm{E}) - \left(\bm{S}_k^\text{v} : \bm{E} - \Phi_k(\bm{S}_k^\text{v})\right) .
	\end{aligned} 
\end{equation}
The symbol $\Psi^\infty$ denotes the free energy of the relaxed elastic solid, such that $\bm{S}^\infty = \partial \Psi^\infty \!/\partial \bm{E}$ is the corresponding Piola--Kirchhoff stress. Introducing the variable
\begin{equation}
	\begin{aligned}
	{\bm Q}_k &= {\partial \Upsilon_k}/{\partial\bm{E}} = g_k \bm{S}^\text{e} - \bm{S}_k^\text{v} \\
	&= g_k \int_0^t \text{e}^{-(t-s)/\tau_k}\, \dot{\bm S}^\text{e}(s) \, \text d s 
	\end{aligned}
\label{Holzapfel}
\end{equation}
yields the constitutive law $\bm{S} = \bm{S}^\infty + \sum_k \bm{Q}_k$ and the rate equation $\dot{\bm Q}_k = g_k \dot{\bm S}^\text{e} -  {\bm Q}_k/\tau_k$.
These expressions are very similar to Sec.~6.10 of the monograph by Holzapfel \cite{holzapfel00}. Indeed, the latter follow from the works of Simo \cite{simo87} and Holzapfel and Simo \cite{holzapfel96b}, as well as Govindjee and Simo \cite{govindjee92}.
Thus, the above nonlinear viscoelasticity theories are strongly related to QLV. Moreover, they are equivalent in the incompressible limit (see \ref{app:Incomp}), as observed in previous work \cite{jridi18,balbi18}.

The expression of the free energy may also be rewritten as a hereditary integral. Miller and Chinzei \cite{miller02} and related studies \cite{taylor09,derooij16} introduce the convolution product $\Psi^* = \mathscr{G} * \dot{W}$ from which the stress $\bm{S} = \mathscr{G} * \dot{\bm S}^\text{e}$ is derived. While the final expression of the stress can easily be related to the present study (Eq.~\eqref{FungConvScalar}), it is not as straightforward to explain the form of the free energy $\Psi^*$. This may be a consequence of a daring differentiation of the free energy with respect to the instantaneous strain inside the convolution integral (see Ref.~\cite{taylor09} Eq.~(8)). Note that in the infinitesimal strain limit, the consistent potential energy reads as a double convolution of the relaxation function with the strain rates (see e.g. Carcione \cite{carcione15} Chap.~2). This remark further questions the expression $\Psi^*$ of the free energy as a single convolution product.

\section{Illustrations}\label{sec:Illustrations}

The theoretical analysis of Sec.~\ref{sec:Thermo} is now illustrated by means of simple deformations. We consider compressible neo-Hookean QLV solids described by the strain energy \cite{ogden84}
\begin{equation}
	W = \tfrac12\mu\, ( {I}_1 - 3 - 2\ln J) + \tfrac12 \mu'\, (J-1)^2 \, ,
\end{equation}
where the Lam{\'e} parameters $\mu'$, $\mu$ are positive. The corresponding bulk modulus reads $\mu' + \frac23 \mu$. The elastic response $\bm{S}^\text{e} = \partial W/\partial \bm{E}$ deduced from Eqs.~\eqref{ElasticSplitTilde}-\eqref{ElasticSplit} reads
\begin{equation}
	\bm{S}^\text{e} = \mu\, (\bm{I} - \bm{C}^{-1}) + \mu' J (J-1)\, \bm{C}^{-1} . 
	\label{ElasticPlot}
\end{equation}
Due to consistency with linear elasticity in the infinitesimal strain limit, this constitutive law is at least locally invertible. Thus, we introduce the complementary energy density $\Phi(\bm{S}^\text{e})$ such that $\bm{E} = \partial \Phi/\partial \bm{S}^\text{e}$.

\begin{figure}
	\centering
	\begin{minipage}{0.45\textwidth}
		\centering
		(a)
		
		\includegraphics{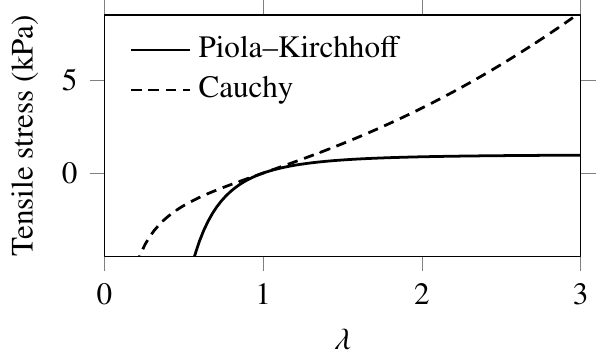}
	\end{minipage}
	
	\begin{minipage}{0.45\textwidth}
		\centering
		(b)
		
		\includegraphics{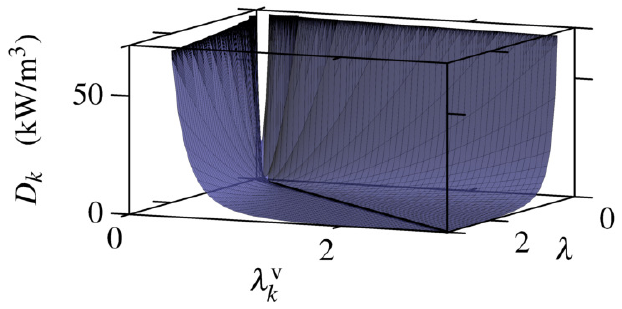}
	\end{minipage}
	
	\caption{(a) Tensile stress of neo-Hookean elastic material with moduli $\mu = 1.0$~kPa and $\mu' = 3\, 000 \,\mu$. (b) Dissipation term of Eq.~\eqref{DissipationTerm}. The black line marks the locus $\lambda = \lambda_k^\text{v}$ of the relaxed elastic solid limit. \label{fig:Compression}}
\end{figure}

The components of $\bm F$ and $\bm{C}$ for uniform extension along the $X$-direction are of the form \cite{ogden84}
\begin{equation}
	\bm{F} = \begin{bmatrix}
		\lambda & 0 & 0 \\
		0 & \sqrt{J/\lambda} & 0 \\
		0 & 0 & \sqrt{J/\lambda}
	\end{bmatrix} , \quad
	\bm{C} = \begin{bmatrix}
		\lambda^2 & 0 & 0 \\
		0 & J/\lambda & 0 \\
		0 & 0 & J/\lambda
	\end{bmatrix} ,
	\label{Compression}
\end{equation}
where $\lambda > 0$ is the tensile stretch.
In \emph{simple tension}, the tractions transverse to the $X$-direction vanish, so that
\begin{equation}
J = \tfrac12\left(1 - \vartheta/\lambda + \sqrt{(1 - \vartheta/\lambda)^2 + 4\vartheta}\right)
\label{SimpleTens}
\end{equation}
with $\vartheta = \mu/\mu'$.
The elastic response of Eq.~\eqref{ElasticPlot} becomes
\begin{equation}
\bm{S}^\text{e} = s^\text{e}\, \bm{e}_1\! \otimes \bm{e}_1
\quad\text{with}\quad
s^\text{e} = \mu\, \big(1 - J/\lambda^3\big) \, .
\label{ElasticPlotScalar}
\end{equation}
Fig.~\ref{fig:Compression}a shows the evolution of $s^\text{e}$ with respect to the stretch $\lambda$, as well as the evolution of the Cauchy stress component $\lambda^2 s^\text{e}/J$. Here, we have chosen $\vartheta = \frac13 \times 10^{-3}$, which is a typical value for nearly-incompressible rubber-like soft solids. One observes that the stress-stretch relationship is one-to-one over the range displayed in the figure. Within this range, we can deduce the stretch $\lambda>0$ from $s^\text{e}$ using Eqs.~\eqref{SimpleTens}-\eqref{ElasticPlotScalar}, e.g. by means of a root-finding algorithm.
Thus, we can retrieve the deformation in Eq.~\eqref{Compression} from the stress.

The viscoelastic stress $\bm S$ is given by the convolution product in Eq.~\eqref{FungConvScalar}. This constitutive law is rewritten as $\bm{S} = \bm{S}^\text{e} - \sum_k \bm{S}^\text{v}_k$ in terms of the memory variables $\bm{S}^\text{v}_k$, which depend on the whole history $t \mapsto \bm{S}^\text{e}(t)$ of the elastic stress. Here, the viscous stresses are of the form $\bm{S}_k^\text{v} =  s_k^\text{v}\, \bm{e}_1\! \otimes \bm{e}_1$ for all times, where
\begin{equation}
	s_k^\text{v} = g_k \int_0^t \big(1-\text{e}^{-(t-s)/\tau_k}\big)\, \dot{s}^\text{e}(s) \, \text d s
	\label{FungMemVarScalar}
\end{equation}
is obtained by componentwise integration of Eq.~\eqref{FungMemVar}.

The viscous strains $\bm{E}_k^\text{v}$ are deduced from the viscous stresses, as specified in Sec.~\ref{sec:Thermo}. Thus, these memory variables are also functions of the entire stress history. In fact, one shows that $\bm{E}_k^\text{v} = \partial \Phi(\bm{S}^\text{v}_k/g_k)/\partial (\bm{S}^\text{v}_k/g_k)$ by using the definition of $\Phi_k$ with respect to $\Phi$. In other words, the mappings $\bm{E}\mapsto \bm{S}^\text{e}$ and $\bm{E}_k^\text{v}\mapsto \bm{S}^\text{v}_k/g_k$ based on the strain energy $W$ are the same, and they admit the same inverse based on the complementary energy $\Phi$. In the present one-dimensional case, it suffices to replace $s^\text{e}$, $\lambda$ by $s_k^\text{v}/g_k$, $\lambda_k^\text{v}$ in Eqs.~\eqref{SimpleTens}-\eqref{ElasticPlotScalar} to retrieve the stretches $\lambda_k^\text{v}$ from the viscous stresses $s_k^\text{v}$. This way, the corresponding deformations $\bm{E}_k^\text{v}$ are obtained. Note in passing that besides being quite involved, the computation of $\bm{E}_k^\text{v}$ is not needed in most practical applications.

The dissipation $\mathscr{D}$ of Eq.~\eqref{PotInternalFinal} involves the sum of terms of the form
\begin{equation}
	\begin{aligned}
	D_k &= \big(\bm{E} - \bm{E}_k^\text{v}\big) : \big(\bm{S}^\text{e} - {\bm{S}_k^\text{v}}/{g_k}\big) \\
	&= \tfrac12\big(\lambda^2 - (\lambda_k^\text{v})^2\big) \big(s^\text{e} - s_k^\text{v}/g_k\big)
	\end{aligned}
	\label{DissipationTerm}
\end{equation}
whose evaluation follows from Eqs.~\eqref{SimpleTens}-\eqref{ElasticPlotScalar} in the present one-dimensional case.
The positivity of $D_k$ for all deformations and all evolutions is illustrated in Fig.~\ref{fig:Compression}b. The black line marks the locus $\bm{E} = \bm{E}_k^\text{v}$ of the relaxed elastic solid limit where $D_k$ vanishes. Finally, the compressible neo-Hookean QLV solid model is thermodynamically admissible in simple tension over the range of stretches considered here.

\section{Conclusion}\label{sec:Conclusion}

In the framework of Fung's quasi-linear viscoelasticity model, the reinterpretation of the natural memory variables as internal variables of state provides an expression of the free energy. The dissipation is shown to be positive provided that the strain energy function is convex. We note that the model equations so-obtained can be linked to other models in the literature.

We should be aware of the limitations of the QLV model, which is known to hardly capture the discrepancy between creep and relaxation time scales \cite{holzapfel17} and which in general does not exhibit strain-dependent relaxation effects. Moreover, QLV may only be valid at moderate deformations \cite{derooij16}. However, its strong similarities to other models mean that its regimes of validity are more-or-less equivalent. The above results could be extended and employed in the modelling of anisotropic materials \cite{balbi18} and thermoelastic materials \cite{holzapfel00}. The results may also be extended to compressible materials with distinct relaxation functions in shear and in compression \cite{depascalis14}. More general results could be obtained by exploiting the notion of fading memory, and the similarity between QLV and the linear viscoelasticity formalism \cite{fabrizio85,fabrizio92}.

\section*{Acknowledgment}

This work was supported by the Irish Research Council under project ID GOIPD/2019/328. Parnell is grateful to the Engineering and Physical Sciences Research Council for funding his Fellowship Extension (EP/S019804/1).

\addcontentsline{toc}{section}{References}

\bibliography{biblio}{}

\appendix

\section{Incompressible case}\label{app:Incomp}

In the case of \emph{incompressible} materials, the volume dilatation
$J \equiv 1$
is prescribed at all times, and the mass density $\rho = \rho_0$ is constant.
The deformation $\bm E$ is equal to its volume-preserving counterpart $\tilde{\bm E} = \frac12 (\tilde{\bm C} - \bm{I})$.
Moreover, the third invariant $I_3 \equiv 1$ is prescribed too, so that the strain energy can be reduced to a function $W(I_1, I_2)$.
The constitutive law \eqref{FungConvScalar} becomes
\begin{equation}
	\bm{S} = -p\bm{C}^{-1}\! +  \mathscr{G} * \dot{\bm S}^\text{e} = -p\bm{C}^{-1}\! + \dot{\mathscr{G}} * {\bm S}^\text{e} ,
	\label{FungConvIncomp}
\end{equation}
where ${\bm S}^\text{e} = \partial W/\partial \bm{E}$ is deduced from Eq.~\eqref{ElasticSplit} with $W_3 = 0$, and $p$ is an indeterminate Lagrange multiplier due to the incompressibility constraint.

In a similar fashion to the compressible case \eqref{FungMemStress}, the constitutive law \eqref{FungConvIncomp} is expressed as
\begin{equation}
	\begin{aligned}
	\bm{S} &= -p\bm{C}^{-1}\! + \bm{S}^\text{e} - \sum_{k=1}^n \bm{S}_k^\text{v}\, ,
	\end{aligned}
	\label{FungStressIncomp}
\end{equation}
where the viscous stresses $\bm{S}_k^\text{v}$ are governed by the linear evolution equation \eqref{FungEvol}.
We define the free energy $\Psi$ using the same expression as in Eq.~\eqref{PotPotStress}.
This way, the Piola--Kirchhoff stress of Eq.~\eqref{FungConvIncomp} satisfies $\bm{S} = -p\bm{C}^{-1} + \partial \Psi/\partial \bm{E}$.
The incompressibility constraint is included in the Clausius--Duhem inequality \eqref{CDT} by introducing the Lagrange multiplier $p$ as follows:
\begin{equation}
	\mathscr{D} = \left(\bm{S} - \frac{\partial \Psi}{\partial \bm{E}} + p\bm{C}^{-1}\right):\dot{\bm E} - \sum_{k=1}^n \frac{\partial \Psi}{\partial \bm{S}_k^\text{v}} : \dot{\bm S}_k^\text{v} \geq 0\, ,
	\label{CDTIncomp}
\end{equation}
see Holzapfel \cite{holzapfel00} Sec. 6.3.
The next steps of the derivation are analogous to the compressible case, and finally, we obtain the same expression of the dissipation as in Eq.~\eqref{PotInternalFinal}. Therefore, the positivity of the dissipation is guaranteed for convex strain energy functions $W$.

\begin{remark}
	In Eq.~\eqref{FungConvIncomp}, the elastic stress $\bm{S}^\text{e}$ may include hydrostatic stress contributions{\,---\,}in other words, we have $\bm{S}^\text{e}:\bm{C} \not\equiv 0$.
	Assuming the elastic stress $\bm{S}^\text{e}$ purely deviatoric is rather restrictive. Indeed, the corresponding Lagrange--Charpit equations yield $W = f(I_1/\!\sqrt{I_2})$ where $f$ is an arbitrary function.
	Sometimes, incompressible QLV is formulated as follows \cite{depascalis14}
	\begin{equation}
		\bm{S} = -p\bm{C}^{-1}\! +  \mathscr{G} * \dot{\bm S}_\text{D}^\text{e} = -p\bm{C}^{-1}\! + \dot{\mathscr{G}} * {\bm S}_\text{D}^\text{e} \, ,
		\label{FungConvIncompDev}
	\end{equation}
	where ${\bm S}_\text{D}^\text{e} = \text{Dev}(\bm{S}^\text{e})$ is purely deviatoric. To proceed as above, one may replace the strain energy $W$ by the strain energy
	\begin{equation}
		W^\star = W(I_1,I_2) - \tfrac16(\bm{S}^\text{e}:\bm{C}) (I_3-1)
	\end{equation}
	in Eq.~\eqref{PotPotStress}, which is defined in such a way that $\bm{S}_\text{D}^\text{e} = \partial W^\star\! /\partial\bm{E}$ under the incompressibility constraint. Alternatively, one may extend $\tilde{W}(\tilde{I}_1,\tilde{I}_2)$ to non-unimodular deformation gradients, such that $\bm{S}_\text{D}^\text{e} = \partial\tilde{W} /\partial\bm{E}$ is satisfied when incompressibility is assumed \cite{pence14}. Thus, the analysis of the dissipation's sign for Eq.~\eqref{FungConvIncompDev} seems more intricate.
\end{remark}

\end{document}